\documentclass[epsf,prl,aps,twocolumn,showpacs,preprintnumbers,amsmath,amssymb,footinbib]{revtex4}
\usepackage{epsf}
\usepackage[colorlinks=true, pdfstartview=FitV, linkcolor=red, citecolor=blue, urlcolor=blue]{hyperref}
\usepackage{bm}
\newcommand{\beq}{\begin{equation}}

\def\arnps#1#2#3{Ann.\ Rev.\ Nucl.\ Part.\ Sci.\  {\bf #1}, #2 (#3)}
\def\atmp#1#2#3{Adv. Theor. Math. Phys. {\bf #1}, #2 (#3)}

\def\ibid#1#2#3{{\it ibid.} {\bf #1}, #2 (#3)}

\def\jhep#1#2#3{Jour. High Energy Phys. {\bf #1}, #2 (#3)}

\def\npa#1#2#3{Nucl. Phys. A {\bf #1}, #2 (#3)}
\def\npb#1#2#3{Nucl. Phys. B {\bf #1}, #2 (#3)}

\def\prc#1#2#3{Phys. Rev. C {\bf #1}, #2 (#3)}
\def\prd#1#2#3{Phys. Rev. D {\bf #1}, #2 (#3)}
\def\prl#1#2#3{Phys. Rev. Lett. {\bf #1}, #2 (#3)}

\def\Tc{T_\text{c}}
\def\Nc{N_\text{c}}
\def\Nf{N_\text{f}}
\def\SU{\text{SU}}

\def\tr{\text{tr}}
\begin{document}
\title{Suppression of the Shear Viscosity in a ``semi'' Quark Gluon Plasma}
\author{Yoshimasa Hidaka,$^{a}$
and Robert D. Pisarski$^{b}$}
\affiliation{
$^a$RIKEN BNL Research Center, Brookhaven National Laboratory, Upton, NY 11973, USA\\
$^b$Department of Physics, Brookhaven National Laboratory, Upton, NY 11973, USA\\
}
\begin{abstract}
We consider QCD at temperatures $T$ near $\Tc$, where the theory deconfines.
We distinguish between a ``complete'' 
Quark Gluon Plasma (QGP), where the vacuum expectation
value of the 
renormalized Polyakov loop is near unity, essentially constant with $T$,
and the ``semi''-QGP, where the loop changes strongly with $T$.
Lattice simulations indicate that in QCD,
there is a semi-QGP from below $\Tc$ to a few times that.
Using a semi-classical model, we compute the shear viscosity,
$\eta$, to leading order in perturbation theory.  We find that
near $\Tc$, where the expectation value of the Polyakov loop is small, that
$\eta/T^3$ is suppressed by two powers of the loop.
For heavy ions, this suggests that
during the initial stages of the collision,
hydrodynamic behavior at the LHC 
will be characterized by a shear viscosity which is significantly larger
than that at RHIC.
\end{abstract}
\date{\today}
\maketitle

The collisions of heavy ions at RHIC have 
demonstrated clear signals for a novel regime.  Much interest has focused
on collective properties, especially elliptical flow, which appear to
be well described by a system in which the 
(dimensionless) ratio of the shear viscosity, 
$\eta$, to
the entropy, $s$, is small 
\cite{teaney,heinz,hirano,romatschke,anomalous,greiner,niemi,csernai,hadronic_viscosity,sqgp}.
These results do not agree with the expectations of a weakly coupled
plasma, and have been described as a 
``strong'' Quark Gluon Plasma (QGP) \cite{hirano,sqgp}.

It is reasonable to expect that 
the running coupling in QCD, $\alpha_s(T)= g^2(T)/(4 \pi)$, 
might be large near the critical temperature,
$\Tc$.  Typically, one expects the coupling to be
in a non-perturbative regime for momenta
less than $\sim 1$~GeV.  Since the transition temperature is 
$\sim 200$~MeV \cite{lattice,loop1,loop2,lattice_viscosity}, 
the coupling could well be large at several (and maybe many) times $\Tc$.

Indeed, perhaps the coupling is so large that the
relevant limit is of infinite coupling.
It is possible to compute
in a ${\cal N} = 4$ supersymmetric $\SU(\Nc)$ theory when the number
of colors, $\Nc$, and $\alpha_s  \Nc$, are both infinite
\cite{susy2,susy3,susy4}.
The ${\cal N} = 4$ supersymmetric theory is conformally invariant, 
so $\eta/s$ is independent
of temperature, and is small, $=1/4 \pi$  \cite{susy2,susy3}.  
If QCD is analogous to the ${\cal N}=4$ theory \cite{susy4}, then
for some region above $\Tc$, 
$\eta/s$ should remain small and not change markedly 
with temperature.

To have a coupling which is relatively moderate in strength at $\Tc$ 
would be exceptional.  At least for the pressure, 
this might occur because of the ubiquitous factors of $2 \pi$ which
accompany the temperature $T$
in the imaginary time formalism \cite{braaten}.  
Using an effective theory in three dimensions \cite{kajantie},
a two loop computation shows that this does, in fact, occur:
at $\Tc$, the effective coupling is only
$\alpha_s^{\rm eff} \approx 0.3$ \cite{laine}.

The challenge is then to understand
how the confining transition, with a large decrease in pressure, occurs
for moderate coupling \cite{rdp}.
We find it useful to view deconfinement as the ionization of
color charge.  Without dynamical quarks, in the confined
phase there is no ionization of color.
Conversely, far into the deconfined phase,
color is completely ionized, either with or without quarks.
In a non-Abelian gauge theory, 
the expectation value of
the (renormalized) Polyakov loop characterizes
the degree to which color charge is ionized.  
Lattice simulations \cite{lattice} find that 
the Polyakov loop is small near $\Tc$, and near one at a few
times $\Tc$ \cite{loop1,loop2}.  
This regime, which we view as one of partial ionization, coincides
with the drop in the pressure (relative to the ideal gas term).
For want of a better term, we refer to this as the ``semi''-QGP.
Above a few times $\Tc$, there is a ``complete'' QGP, where
both the renormalized Polyakov loop, and the pressure$/T^4$,
are essentially constant.

In this paper we consider how the shear viscosity changes in the semi-QGP.
We make numerous drastic assumptions.  If the 
coupling is moderate even down to $\Tc$, perhaps
we can treat the semi-QGP by means of 
a semi-classical approximation.  Our ansatz is extremely simple,
just a background field with constant $A_0$, Eq. (\ref{ansatz}).  We then 
compute to leading order in $\alpha_s$ 
and $\log(1/\alpha_s)$ \cite{lebellac,amy},
for an infinite number of colors and flavors \cite{planar}.
We ignore the change of $\eta$ with $\alpha_s$, to
concentrate on how it changes as the 
expectation value of the Polyakov loop decreases.
We find that when the loop is small, that the shear viscosity is suppressed
by two powers of the Polyakov loop.
This implies that $\eta/T^3$
decreases significantly in the semi-QGP
as the theory cools, from a few times $\Tc$ down to $\Tc$.
Since $\eta \sim 1/\alpha_s^2$ at small $\alpha_s$, Eq. (\ref{eq_viscosity}),
including the running of the QCD coupling could
reinforce this trend.  
Such a large decrease near $\Tc$ is very different from
the strong QGP, where $\eta/s$ changes little \cite{hirano,sqgp}, 
if at all \cite{susy4}, with temperature.

There are numerous examples of non-relativistic systems which exhibit
a minimum in the shear viscosity near $\Tc$ 
\cite{csernai,susy2}.
The present analysis suggests how this might arise dynamically in QCD;
see, {\it e.g.} \cite{hirano}.
Our result is also reminiscent of the ``anomalous'' viscosity
of an Abelian plasma, where the viscosity changes because of a background
field, although our specific
background field is not like that of Abelian plasmas \cite{anomalous}.

The Polyakov loop represents the propagator of an
infinitely massive test quark, which we assume is in the fundamental
representation 
\cite{loop1,loop2,rdp,interface,aharony,gw,dumitru}.  
Its magnitude can be viewed as the probability for the test quark to
propagate.  This is near one at asymptotically high temperature, 
where the plasma is completely ionized.

In a $\SU(\Nc)$ gauge theory without dynamical quarks, in the confined
phase the propagator for a test quark vanishes identically.
This is because the Polyakov loop carries
Z(${ \Nc}$) charge, and the confined phase is
Z(${ \Nc}$) symmetric.  Thus without dynamical quarks, there is 
absolutely no ionization of Z(${ \Nc}$) charge below $\Tc$, only above.

The expectation value of the renormalized Polyakov 
loop, $\langle \ell \rangle$, 
is extracted from that of the bare loop after 
a type of mass renormalization \cite{loop1}.
For a $\SU(3)$ gauge theory without quarks, 
from Fig. (1) of Ref. \cite{loop1},
$\langle \ell \rangle = 0$ below $\Tc$;
at $\Tc^+$, $\langle \ell \rangle \sim 0.5$; it crosses one at $\sim
2.8 \, \Tc$, and reaches $\sim 1.1$ by $\sim 4 \, \Tc$.
This value is then constant
from $\sim 4 \, \Tc$ up to to highest temperature measured, $\sim 12 \, \Tc$.
(In perturbation theory, $\langle \ell \rangle$ exceeds unity.)
Thus in a $\SU(3)$ gauge theory without quarks,
there is a confined phase below $\Tc$, 
a semi-QGP from (exactly) $\Tc^+$ to about $\sim 4 \, \Tc$,
and a complete QGP above that.
By its nature, the boundary between the semi- and the complete QGP
is not precise.

In QCD the Polyakov loop is no longer a
strict order parameter, since dynamical quarks
also carry Z(${ \Nc}$) charge.  
Thus even below $\Tc$, the 
Z(${ \Nc}$) charge of a test quark is shielded by the thermal
ionization of dynamical quark anti-quark pairs.
With sufficiently many flavors of quarks, this could happen
even at rather low temperature.

At present, however, numerical simulations indicate only a modest
ionization of color below $\Tc$,
at least for three colors and $2+1$ flavors of dynamical quarks.
From Fig. 11 of Ref. \cite{loop2}, the expectation value of the 
Polyakov loop is very small below 
$0.8 \, \Tc$; it then rises to
$\sim 0.3$ at $\Tc$, and is near one at $\sim 2 \, \Tc$.
We take this to show that in QCD that there is a semi-QGP
between $\sim .8 \, \Tc$ to perhaps $\sim 2-3 \, \Tc$.  
Thus with dynamical quarks, there is a semi-QGP in 
both the hadronic phase, from $\sim 0.8 \, \Tc$ to $\Tc$, 
and in the deconfined phase, from $\Tc$ to $\sim 2-3 \, \Tc$;
again, there is a complete QGP above that.

We characterize the semi-QGP by the following approximation
\cite{rdp,interface,aharony,gw,dumitru}.
The Polyakov loop is the trace of a straight Wilson line in imaginary
time.  An expectation 
value for the Polyakov loop which is not near one implies
a non-trivial distribution for the eigenvalues of this Wilson line.
We thus expand 
about a constant, background field for the timelike component of the
vector potential, 
\begin{equation}
A_0^\text{cl} = Q/g \; .
\label{ansatz}
\end{equation}
where $g$ is the coupling constant 
for an $\SU(\Nc)$ gauge theory.  The matrix $Q$ is diagonal in color
space, with the indices those for the fundamental representation,
$a = 1...{ \Nc}$.  The Wilson line is ${\rm L} = \exp(i Q/T)$, and 
the bare Polyakov loop is
$\ell = \tr \, {\rm L}/ \Nc$.
Although we expand about a given $Q$, what is physically relevant is the
distribution of $Q$'s.  The moments of this distribution are
given by powers of the Wilson line, 
$ \tr \, {\rm L}^j/  \Nc$, $j= 1 \ldots ( \Nc-1)$.

In a complete theory we would self-consistently determine
the effective Lagrangian which gives the distribution of the $Q$'s.
Notably, this would enable one to compute the pressure (and thus
the entropy) as a function
of the $Q$'s; e.g., versus $\langle \ell \rangle$, etc.
Some results for the $Q$-distribution can be obtained analytically when
the spatial volume is a small sphere \cite{aharony}.  
In a large volume, its form is unknown \cite{rdp,gw};
its determination through
numerical simulations of effective theories appears promising 
\cite{dumitru}.  In lieu of this, 
we take the first moment of the Wilson line
from lattice simulations in the full theory \cite{loop1,loop2},
and choose two different forms for the complete eigenvalue distribution.
We find, unexpectedly, that at least the shear viscosity is rather insensitive
to which distribution we take.

The shear viscosity is computed using a Boltzmann 
equation \cite{lebellac,amy} in the presence of the background
field, $Q$.  In the imaginary time formalism, the Euclidean
four momenta are 
$P_\mu = (p_0,\bm{p})$, where $p_0$ is an even (odd) multiple
of $\pi T$ for bosons (fermions).  To compute with $Q\neq 0$
we use the double line notation of 't Hooft
\cite{planar}.  To simplify the calculation we take $\Nc$ and $\Nf$ to be
infinite, although the method can be used at finite values \cite{tocome}.
How the $Q$'s enter depends upon the representation of the gauge group.  
Quarks in the fundamental representation have one color line, so 
the four momenta has one color index,
$P^a_\mu = (p_0 + Q^a,\bm{p})$, where $Q^a$ is the diagonal component 
of the matrix $Q$  (we also use $P^{-a}_0 = p_0 - Q^a$).
Adjoint gluons carry two color lines, and their momenta has two color indices,
$P^{a b}_\mu = (p_0 + Q^{a b},\bm{p})$, defining
$Q^{a b} = Q^a - Q^b$.
Each component $Q^a$ is typically
a non-integral multiple of $2 \pi T$, and so $Q^a \neq 0$
cannot be shifted away without violating 
the appropriate boundary conditions in 
imaginary time \cite{lebellac}.  
$Q^a$ is like an imaginary chemical potential
for color charge, in the space of 
diagonal (mutually commuting) generators.  

Amplitudes are analytically continued from imaginary to real time
in the usual fashion, except that now it is necessary to analytically
continue using energies which carry color, $p_0 + Q^a \rightarrow -i\omega^a$
for quarks, and $p_0 + Q^{a b} \rightarrow -i \omega^{a b}$ for gluons
\cite{furuuchi}.  The mass shells remain on the light cone, 
$\omega = \pm E$, $E = \sqrt{\bm{p}^2}$.

For quarks, what typically enters are distribution functions
$\widetilde{n}(E - i Q^a)$, 
where $\widetilde{n}(E)$ 
is the usual Fermi-Dirac statistical distribution function.
To compute, expand this as \cite{aharony}
\begin{equation}
\frac{1}{{\rm e}^{(E - i Q^a)/T} + 1}
= \sum_{j=1}^{\infty} (-)^{j+1} {\rm e}^{- j(E - i Q^a)/T} \; .
\label{expansion}
\end{equation}
The first term, $\sim \exp(- E/T)$, 
represents the Boltzmann approximation to the quantum
distribution function, and
is accompanied by $\exp(i Q^a/T)$.
Expansion to $j^\text{th}$ order brings in a factor of $\sim \exp(- j E/T)$, 
and is accompanied by $\exp(i j Q^a/T)$.  
For a given process, the moments of the Wilson line
which enter depend upon the detailed routing of
the color indices and the like.  As an example,
consider the trace of the quark propagator:
then the first, Boltzmann term involves the
trace of $\exp(i Q^a/T)$, which is the Polyakov loop, $\ell$;
terms to $j^\text{th}$
order involve the $j^\text{th}$ moment of the Wilson line,
${\tr} \, {\rm L}^j$.

The gluon propagator is similar,
except that 
$n(E - iQ^{a b})$, enters, where $n(E)$ is the Bose-Einstein statistical
distribution function.  Again, as an example
consider summing over the indices of the gluon propagator.
To avoid taking the trace, which is part of the correction
in $1/ \Nc$, one sums separately over $a$ and $b$.
The first, Boltzmann term
involves the traces of $\exp(i Q^{a b}/T)$, which becomes
$| \tr \, \rm L|^2$; terms to
$j^\text{th}$ order become $|{\tr} \, {\rm L}^j |^2$.

In computing perturbatively
about a trivial vacuum, one naturally divides the momenta into hard momenta,
whose components are $\sim T$, or soft momenta, where both the energy,
$\omega$, and spatial momenta, $p$, are $\sim gT$ \cite{lebellac}.  
This remains valid at $Q^a \neq 0$ for the colored energies,
$\omega^a$ and $\omega^{a b}$; the $Q$'s themselves are $\sim T$, and so
hard.

We have computed the hard thermal loops (HTLs) in the quark and gluon self
energies for $Q \neq 0$ \cite{tocome}.  HTL's are the dominant contributions
when the external momenta are soft.  To leading order, $\sim g^2$,
the dominant term involves an
integral over hard momenta, times an angular integral over soft momenta
\cite{lebellac}.  The dependence on the $Q$'s only affect the integral 
over the hard momenta, through the change in the statistical distribution
functions; the angular integral is unchanged.

For the quark self energy, the only change in the HTL is the change in
the Debye mass, which now depends upon the direction in color space.
For the gluon self energy, there is the usual HTL, again modified by
the change in the Debye mass.  Besides the HTL, which is $\sim g^2 T^2$,
there are also terms $\sim g^2 T^3 Q/\omega$.  These new terms
are directly proportional to tadpole terms, and arise because 
$Q \neq 0$ induces a net color charge.
To understand the terms $\sim T^3$, consider a Z($\Nc$) interface,
which is modeled by a spatially dependent
$Q$ \cite{interface}.  There such terms are physical, because the
interface induces a nonzero color electric field \cite{tocome}.
For the present problem, however, we assume that there is no net color
charge in the vacuum, so the only change in the soft gluon 
propagator is the change of the Debye mass.  
(If the gluon HTL does include a term $\sim T^3$, 
then for processes involving exchange of a soft gluon,
in Eq. (\ref{eq_viscosity}) $\log(1/g) \rightarrow \log(1/g^{2/3})$).

We now outline the calculation of the shear viscosity \cite{tocome}.
To leading order, both in $g^2$ and in $\log(1/g)$,
we find that the result can be written as \cite{lebellac,amy,tocome}
\begin{equation}
\frac{\eta}{T^3} = \frac{c_\eta}{g^4 \; \log(1/g) } \; f({\rm L}) \; .
\label{eq_viscosity}
\end{equation}
The constant $c_\eta$ depends upon the number of colors,
$\Nc$, and flavors, $\Nf$ 
\cite{amy}.  As noted before, for calculational reasons we 
compute for $\Nc = \infty$, with $ \Nf/ \Nc$ fixed, implicitly assuming
that $f({\rm L})$ is relatively insensitive to $\Nc$.

At this order, the 
viscosity is determined by the scattering of $2 \rightarrow 2$
particles, where all particles have hard momenta.  They interact through
the exchange of a single, soft field in the $t$-channel.
For the soft field, we use the HTL approximation.
The Debye mass changes when $\rm L \neq 1$, but this doesn't enter at
leading logarithmic order.

For the pure glue theory at infinite $\Nc$, to leading 
order in $g^2$ and $\log(1/g)$ \cite{lebellac,amy},
two scattering processes contribute to the collision term:
$P_1^{a b} + P_3^{b c} \rightarrow P_2^{a d} + P_4^{d c}$
and
$P_1^{a b} + P_3^{c d} \rightarrow P_2^{a d} + P_4^{c b}$.
This is nearly forward scattering, with 
the spatial momenta $\bm{p}_1 \approx \bm{p}_2$, and
$\bm{p}_3 \approx \bm{p}_4$.  The momenta of the exchanged gluon,
$\bm{p}_1 - \bm{p}_2 =
\bm{p}_4 - \bm{p}_3$, must be soft to give $\log(1/g)$.

\begin{figure}
\begin{center}
\epsfxsize=.40\textwidth
\epsfbox{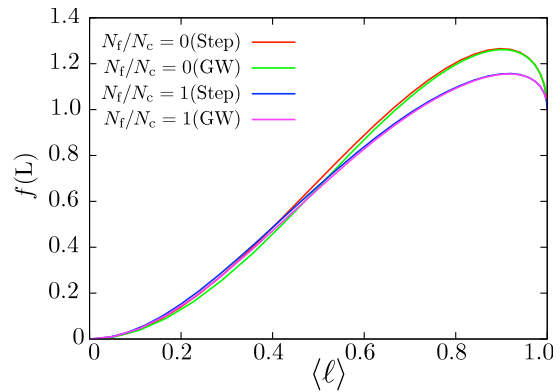}
\end{center}
\caption{The function $f(\rm L)$ of Eq. (\ref{eq_viscosity}), versus $\ell$.
`Step' and `GW' denote  $Q$ distributions with a simple step-function
and  that in the Gross-Witten matrix model \cite{aharony,gw}, respectively.
}
\label{viscosity1}
\end{figure}

Under these approximations, the
dependence upon the background field $Q$ only enters through the
integrals over the statistical distribution functions for the hard
fields, $p_1$ and $p_2$.  While the statistical distribution functions
are complex when $Q \neq 0$, 
after summing over both emission and absorption processes,
all contributions to the stress energy tensor are real.
For the shear viscosity, the integrals which enter are
\begin{eqnarray}
&&\hspace{1.5cm} \int^\infty_0 dp \; p^4 \; n\!\left(p - iQ^{a b}\right) \; , \notag\\
&&\int^\infty_0 dp \; p^j \; n\!\left(p - i Q^{a b} \right) 
\left(1 + n\!\left(p - i Q^{a d} \right) \right) \; ,
\label{integral}
\end{eqnarray}
where $j = 2$ and $4$.  One then expands the distribution functions as
in Eq. 
(\ref{expansion}), to obtain a power series in moments of the Wilson line.
With quarks, there are more diagrams, as the distribution functions
between quarks and gluons mix \cite{amy}.
The integrals are like Eq. (\ref{integral}), but also involve 
$\widetilde{n}(E - i Q^a)$.
(We also take $\chi(p)\sim p^2$ \cite{amy}: this is valid to 
to $< 1.0\%$ for $Q=0$, and is exact at small $\ell$, where
Boltzmann statistics applies.)

As written in Eq. (\ref{eq_viscosity}), the result is a
function of the Wilson line, $f(\rm L)$, times the usual perturbative result;
thus $f=1$ when $\rm L = 1$.  This simple form is not valid beyond leading
logarithmic order.  If $\ell = \tr \, {\rm L}/ \Nc$ 
is small, and dominates higher moments, we find the result vanishes like
the square of the loop:
\begin{equation}
f(\text{L}) \approx
a_2(\lambda) \; \left(
\frac{\ell+4 \lambda}{\ell+\lambda} 
 \right) 
\ell^2 \; , \; \ell \ll 1  \;,
\label{smallL}
\end{equation}
where $\lambda =  \Nf/ \Nc$:
$a_2(0)\approx 3.31$, $a_2(1) \approx 1.01$.

The behavior at small $\ell$ can be understood as follows.  
In the pure glue theory, when gluons with momenta
$P_1^{a b}$ and $P_3^{b c}$ scatter,
the $Q^b$ charges cancel,
so summation over $a$ and $c$ gives a collision term $\sim g^4 \ell^2$ 
(times $\log(1/g)$, which we suppress for brevity).
There is no such cancellation for the scattering of gluons with momenta
$P_1^{a b}$ and $P_3^{c d}$, which is only $\sim g^4 \ell^4$.
The source term \cite{amy} is like
the trace of the gluon propagator, $\sim \ell^2$, and so without
quarks, at small $\ell$
the gluon contribution to the shear viscosity is
$\eta_\text{gluon} \sim (\ell^2)^2/(g^4 \ell^2) \sim \ell^2/g^4$.  

With quarks, the collision term is dominated by a 
quark scattering off of an anti-quark,
$P_1^{a} + P_3^{-a} \rightarrow P_2^{c} + P_4^{-c}$.
The $Q^a$ charges  cancel, so this is $\sim g^4 \ell^0$ at
small $\ell$.  
If a quark scatters off of a quark,
$P_1^{a} + P_3^{b} \rightarrow P_2^{b} + P_4^{a}$,
the $Q$ charges don't cancel, and the scattering is $\sim g^4 \ell^2$.
Gluons also scatter off of quarks, by exchanging a gluon:
$P_1^{ab} + P_3^{b} \rightarrow P_2^{ac} + P_4^{c}$;
the $Q^b$ charges cancel, so this is $\sim g^4 \ell$.
The mixing of the gluon and quark
distribution functions \cite{amy} can be neglected
at small $\ell$: quark anti-quark annhilation,
$P_1^{a} + P_3^{-b} \rightarrow P_2^{ac} + P_4^{c(-b)}$,
is $\sim g^4 \ell^2$, while Compton scattering, 
$P_1^{a} + P_3^{bc} \rightarrow P_2^{ab} + P_4^{c}$, is $\sim g^4 \ell^3$.
For the source term, the quark contribution
is like summing over
the quark propagator, $\sim \ell$, so the quarks contribute to
the viscosity as 
$\eta_\text{quark} \sim (\ell)^2/(g^4 \ell^0)\sim \ell^2/g^4$.  
In the presence of quarks, gluons
contribute to the viscosity as
$\eta_\text{gluon}\sim (\ell^2)^2/(g^4 \ell) \sim \ell^3/g^4$, down
by $\sim \ell$ to the quark contribution.  

To obtain results valid for all values of $\ell$, some assumption about
higher moments must be made.  We used two forms.
The first is to take a simple step function, of width $\beta$, about
the origin, so that $\tr \, {\rm L}^n/ \Nc = \sin(n \beta)/(n \beta)$.
The other is to take a $Q$ distribution as in the Gross-Witten matrix
model \cite{aharony,gw}.   
By the methods described above, it is straightforward to obtain
results, although their analytic form is unwieldy.  These forms
can be easily evaluated numerically, though, 
as shown in Fig.(\ref{viscosity1}).
We find that $f(\rm L)$ is insensitive to the assumption about
higher moments, changing by at most 
a few percent over the entire range of $\ell$.
We find that with both eigenvalue distributions, that
the collision term generates a cusp in $f(\rm L)$ near $\ell = 1$.
We expect that this non-analytic behavior will be washed out 
by corrections to higher order, which enter for $Q \sim g T$.

We conclude with some general comments.
The Polyakov loop is proportional to the propagator of an infinitely 
massive test quark,
and as such, has no direct relation to the propagation of 
dynamical fields.  In our ansatz, however, the
quasiparticles are fluctuations about the background field
in Eq. (\ref{ansatz}).  When the expectation value of the Polyakov loop
is small, this background field universally
suppresses the propagation of any colored field.
For heavy fields, this is reasonable, but for light
fields, it is nontrivial.   If the light fields have
hard momenta $p \geq T$, then suppression by the Polyakov loop
is the dominant effect, with other corrections down
by powers of $g$.  At soft momenta, $p \sim gT$, light fields are not only
suppressed by a (small) Polyakov loop, but altered by the change in
their hard thermal loops.

In general, the shear viscosity is proportional
to the ratio of a source and a collision term 
\cite{lebellac,amy}.  In the absence of a background field,
the source term is of order one, and so the shear viscosity can only
be small if the coupling constant, and so the collision term, are large. 
This is the central idea which motivates the strong QGP \cite{hirano,sqgp}.
In contrast, in our analysis of the semi-QGP,
the quasiparticles are weakly coupled, but 
are fluctuations about a nontrivial background field.  
It is this background field which
suppresses both the source and collision terms, 
to give $\eta/T^3 \sim |\ell|^2$ at small $\ell$.

We suggest that this is not an artifact of our approximations.
If deconfinement truly represents the ionization of color charge, 
then it is reasonable to expect that 
the propagation of all colored fields are suppressed at small $\ell$.
In particular, while we have 
computed only to leading order in $g^2$ and $1/\log(1/g)$,
one can show that within our ansatz, 
that $\eta/T^3 \sim \ell^2$ as $\ell\rightarrow 0$,
order by order in perturbation theory at $\Nc = \infty$
\cite{tocome}.
The suppression of $\eta/T^3$ near $\Tc$
is testable through simulations on the lattice \cite{lattice_viscosity}.

If color is only
partially ionized in the semi-QGP, then some
color singlet states persist for a range of temperatures above $\Tc$,
before they eventually disassociate into colored constituents.
Similarly, with dynamical quarks colored fields
contribute below $\Tc$, in the hadronic part of the semi-QGP.
Including both effects is clearly challenging; here we just speculate
about the complete result.
Although hydrodynamics depends upon $\eta/s$,
without a complete theory we can only discuss
the behavior of $\eta/T^3$.
Working up from low temperatures in the hadronic phase,
$\eta/T^3$ appears to decrease with increasing $T$, as in a liquid;
{\it e.g.}, for massless pions,
to leading order in chiral perturbation theory
\cite{hirano,hadronic_viscosity,sqgp}.
Working down from high temperatures,
from the complete to the semi-QGP, we have shown that 
the contribution of quarks and gluons to the shear viscosity decreases
as $\eta/T^3 \sim |\ell|^2/\alpha_s^2$
(this assumes that higher order corrections in $\alpha_s$ do not alter the
leading behavior, at least qualitatively).
Combining these two effects, we obtain a minimum for $\eta/T^3$ near
$\Tc$, as in non-relativistic systems \cite{csernai,susy2}.  
It seems implausible that the minimum is precisely at $\Tc$;
certainly in QCD, where there is
no true phase transition \cite{lattice,loop2}.

It is natural to suspect that
heavy ion collisions at RHIC have probed some region in the semi-QGP.
Since one needs a small value of the shear viscosity to fit the
experimental data
\cite{teaney,heinz,hirano,romatschke,anomalous,greiner,niemi,csernai,hadronic_viscosity,sqgp}, perhaps one is near $\Tc$.  
Heavy ion collisions at the
LHC may probe temperatures which are significantly higher, possibly well
into the complete QGP.  If so, then at small times collisions at the LHC
create a system with 
large shear viscosity; as the system cools through $\Tc$, the shear
viscosity then drops.
Thus the semi-QGP predicts that at short times, the hydrodynamic behavior of
heavy ion collisions at the LHC is qualitatively
different from that at RHIC.  This, and other phenomenological implications
of the common suppression of colored fields at small $\ell$, will be
developed separately \cite{tocome}.

In contrast, models of a strong QGP predict
that hydrodynamics at the LHC will be similar to that at RHIC,
characterized by a small value of $\eta/s$
\cite{hirano,niemi,sqgp}.  In particular, while in
${\cal N} = 4$ supersymmetry the pressure$/T^4$ is constant \cite{susy3},
several models have been proposed to fit
the QCD pressure right down to $\Tc$ \cite{susy4}.  
Even so, in all of these models 
$\eta/s$ is independent of temperature, and so remains small \cite{susy4}.  

We eagerly await the
experimental results for heavy ions from the LHC, which 
may be as unexpected and exciting as those from RHIC first were.

This
manuscript has been authorized under Contract No. DE-AC02-98CH10886 with
the U. S. Department of Energy.

\end{document}